\documentclass[trackchanges, twocolumn]{aastex701}

\usepackage[dvipsnames]{xcolor}
\usepackage{amsmath}
\usepackage{graphicx}
\usepackage{siunitx}
\usepackage{placeins}
\usepackage[dvipsnames]{xcolor}

\shortauthors{Zafuta, Havens, et al.}

\begin{document}

\title{Constraining Dust Attenuation Curves in Star-Forming Galaxies at Cosmic Noon Using Multiple Emission Line Ratios}

\author[0009-0003-6859-2292]{Owen Zafuta}
\thanks{These authors contributed equally to this work.}
\affil{Department of Physics and Astronomy, University of Kansas, Lawrence, KS 66045, USA}
\email{ogzafuta@ku.edu}

\author[0009-0006-9141-4381]{Joseph Havens}
\thanks{These authors contributed equally to this work.}
\affil{Department of Physics and Astronomy, University of Kansas, Lawrence, KS 66045, USA}
\email{joe.havens79@ku.edu}

\author[0000-0001-8534-7502]{Bren E. Backhaus}
\affil{Department of Physics and Astronomy, University of Kansas, Lawrence, KS 66045, USA}
\email{bren.backhaus@ku.edu}

\author[0000-0002-5537-8110]{Allison Kirkpatrick}
\affil{Department of Physics and Astronomy, University of Kansas, Lawrence, KS 66045, USA}
\email{akirkpatrick@ku.edu}

\author[0000-0001-5930-0532]{Gregory Troiani}
\affil{Department of Physics and Astronomy, University of Kansas, Lawrence, KS 66045, USA}
\email{gtroiani@ku.edu}


\author[0000-0002-6292-4589]{Kurt Hamblin}
\affiliation{Department of Physics and Astronomy, University of Kansas, Lawrence, KS 66045, USA}
\email{kurt.hamblin@ku.edu}

\begin{abstract}
Dust attenuation remains a dominant systematic uncertainty in measurements of an emission line, and therefore galaxies properties derived from the emission lines such as star formation rates. Corrections based only on the Balmer decrement require an assumed attenuation-curve slope, even though that slope varies among galaxies. We use JWST/NIRSpec spectroscopy from JADES to constrain effective dust attenuation curves for 386 star-forming galaxies at $0.5\leq z\leq7$. For each galaxy, we combine H$\alpha$, H$\beta$, and one additional hydrogen recombination line and solve for the normalization $\tau_V$ and slope $n$ of a power-law attenuation curve, allowing the slope to vary from galaxy to galaxy. We found slopes and normalization span a broad range even for galaxies at similar redshifts.  Additionally, our method using Paschen transitions provided substantially greater constraints the attenuation-curve slope than Balmer-only combinations.
When comparing to SMC-like or Calzetti-like corrections we find our method produces more consistent star formation rates with less scatter.
These improved dust corrections provide an avenue to better describe star formation rates in the epoch of peak star formation and the early universe.
\end{abstract}

\keywords{\uat{Interstellar dust}{836} --- \uat{Galaxy evolution}{594} --- \uat{Emission line galaxies}{507} --- \uat{High-redshift galaxies}{734} --- \uat{Star formation}{1569}}

\section{Introduction}
\label{sec:intro}

Understanding how galaxies form and evolve requires reliable measurements of their star formation rates (SFRs). Hydrogen recombination lines  
trace the ionizing radiation produced by short-lived massive stars and therefore probe recent star formation over timescales of approximately $5$--$10$ Myr \citep{Kennicutt1998, Murphy2011, Kennicutt2012}. Their observed luminosities, however, can be strongly suppressed by interstellar dust, making attenuation corrections one of the dominant systematic uncertainties in nebular SFR measurements \citep[e.g.][]{Vale_Asari_2020}.

The most widely used nebular attenuation \footnote{To describe the combined impact of emission, absorption, and scatter from dust, we employ the term attenuation} diagnostic is the Balmer decrement \citep[e.g.][]{Galliano_Galametz_Jones_2018}, the observed flux ratio of two hydrogen recombination lines, H$\alpha$/H$\beta$. Under Case B recombination, the intrinsic ratios of hydrogen lines are set by atomic physics and depend only weakly on the electron temperature and density of the emitting gas.
An observed Balmer decrement larger than this value indicates stronger attenuation at H$\beta$ than at H$\alpha$.

A single Balmer decrement constrains the relative attenuation between two optical wavelengths but cannot independently determine both the overall attenuation and its wavelength dependence. Conventional corrections therefore assume a fixed attenuation or extinction curve, such as a Calzetti-like or Small Magellanic Cloud (SMC) prescription \citep{Calzetti2001,Gordon2003}. This assumption can be limiting because effective attenuation curves vary among galaxies as a result of differences in dust properties, metallicity, and the spatial distributions of stars, gas, and dust \citep[e.g.,][]{Charlot2000, Calzetti2001, Reddy2015, Salim_2020,Forrey2025}. However, when these assumed slopes are extrapolated out to longer, near-infrared wavelengths, they predict drastically different amounts of dust obscuration.

Hydrogen Paschen transitions provide the additional wavelength leverage needed to test the assumed curve shape. Pa$\gamma$, Pa$\beta$, and Pa$\alpha$ occur at longer wavelengths than the Balmer lines and are therefore less strongly attenuated \citep[e.g.,][]{Calzetti2007,Cleri2022}. Combining H$\alpha$ and H$\beta$ with a third hydrogen line, particularly a Paschen transition, supplies two independent line ratios spanning a larger range of wavelength. 
These ratios can jointly constrain the normalization and slope of a parameterized attenuation curve rather than requiring the slope to be fixed in advance.

Previous studies have demonstrated both the value and the observational difficulty of this approach. \citet{Reddy2015} showed that the average attenuation behavior of $1.4<z<2.6$ star-forming galaxies can differ from canonical local prescriptions. \citet{Prescott2022} developed a three-line framework that uses HST and ground-base Balmer and Paschen measurements from of eleven $z \approx 0.2$ galaxies  to solve simultaneously for attenuation normalization and slope, but their small sample and heterogeneous observing configurations limited the number of well-constrained solutions. Four galaxies yielded physically plausible analytic solutions, although only one retained a well-constrained slope and normalization after observational uncertainties were incorporated. Their study demonstrated that three hydrogen lines can, in principle, constrain both attenuation normalization and slope, but it also showed that aperture differences and slit-loss corrections between separately observed Balmer and Paschen lines can strongly bias the inferred parameters.

Near-Infrared Spectrograph (NIRSpec) observations from \textit{James Webb Space Telescope} (JWST) now enables Balmer and Paschen transitions to be measured within a more uniform spectroscopic data set for substantially larger samples of distant galaxies \citep[e.g.,][]{Cameron2023, Shapley2023, Sandles2024}.

Applying a recombination-line attenuation model also requires controlling contamination from active galactic nuclei (AGN). In composite systems, dust exists at a variety of temperatures and traces structurally distinct regions, ranging from the circumnuclear torus to the broader host-galaxy environment \citep{Kirkpatrick2012}. Consequently, the measured hydrogen emission can include contributions from broad-line, narrow-line, and star-forming regions that do not share a single attenuation curve or a common set of intrinsic Case B ratios \citep[e.g.,][]{Antonucci1993, Urry1995, Veilleux1995, Netzer2015, Hickox2018}. Optical strong-line diagnostics also become increasingly degenerate between stellar and AGN ionizing sources under the high-ionization and low-metallicity conditions common at high redshift \citep{cleri2025}. 
Constraining attenuation in the interstellar medium therefore requires the AGN to be separated out from the star formation, or removed from galaxy samples altogether.


In Section~\ref{sec:data}, we describe the spectroscopic data, emission-line and measurements. In Section~\ref{sec:sample_selection}, we present the sample-selection criteria and our AGN removal methodology. In Section~\ref{sec:analysis}, we introduce the multi-line attenuation framework and present the resulting attenuation parameters. In Section~\ref{sec:SFR}, we compare the SFR corrections produced by the galaxy-specific and fixed-slope attenuation prescriptions. We summarize our principal results in Section~\ref{sec:summary}. We assume a flat $\Lambda$CDM cosmology with $\Omega_{\mathrm M}=0.3$, $\Omega_{\Lambda}=0.7$, and $H_0=70\,\mathrm{km\,s^{-1}\,Mpc^{-1}}$ \citep{Spergel2003}.

\section{Data and Emission-Line Measurements}
\label{sec:data}

\subsection{Parent Surveys} \label{sec:parent_sample}

The parent spectroscopic sample for this study is drawn from observations obtained with the NIRSpec aboard the JWST  \citep{2022A&A...661A..80J} as part of the JWST Advanced Deep Extragalactic Survey (JADES; JWST GTO programs 1180, 1181, 1210, and 1286; \citealt{DEugenio2024}). The JADES observing strategy and target selection 
are described by \citet{DEugenio2024}.

The JADES observations cover the GOODS-North and GOODS-South extragalactic legacy fields. They include the medium-resolution G140M/F070LP, G235M/F170LP, and G395M/F290LP grating--filter configurations, high-resolution G395H/F290LP observations for a subset of targets, and low-resolution PRISM/CLEAR spectroscopy. Together, these configurations provide broad wavelength coverage that enables simultaneous measurements of the Balmer and Paschen transitions required for our multi-line dust attenuation analysis.

We retrieve the uniformly reduced one-dimensional NIRSpec spectra and associated spectroscopic-redshift metadata from version 4 of the DAWN JWST Archive (DJA)\footnote{\url{https://dawn-cph.github.io/dja/spectroscopy/nirspec/}}. The extraction and reduction methodology underlying the DJA NIRSpec data products is described by \citet{deGraaff2024} and \citet{Heintz2023}. Combining the GOODS-North and GOODS-South fields reduces the influence of field-to-field variance and provides a large spectroscopic sample.

\subsection{Emission Line Identification and Measurement}
\label{sec:line_measurements}

We search the rest-frame spectra of 3,156 unique galaxies for hydrogen recombination and nebular emission lines. The extraction pipeline targets H$\delta$, H$\gamma$, H$\beta$, H$\alpha$, Pa$\gamma$, Pa$\beta$, and Pa$\alpha$. The pipeline also measures Ly$\alpha$, Ly$\beta$, Br$\alpha$, Br$\beta$, [O\,II] $\lambda\lambda3726,3729$, [Ne\,III] $\lambda3869$, [O\,III] $\lambda5007$, and [N\,II] $\lambda6583$. 
Table~\ref{tab:emission_lines} lists the rest-frame line centers adopted by the extraction pipeline and the theoretical Case B emissivity ratios relative to H$\alpha$.

\begin{table}[t]
\centering
\caption{Emission Lines Used in the Attenuation and AGN Analyses}
\label{tab:emission_lines}
\begin{tabular}{lccc}
\hline
Ion / Line & Transition & $\lambda_{\mathrm{rest}}$ (\AA) & $j_{\lambda}/j_{\mathrm{H}\alpha}$ \\
\hline
\multicolumn{4}{c}{\textbf{Hydrogen Lines Used in the Attenuation Analysis}} \\
\hline
H$\delta$  & $6\rightarrow2$ & 4102.0  & 0.091 \\
H$\gamma$  & $5\rightarrow2$ & 4341.0  & 0.164 \\
H$\beta$   & $4\rightarrow2$ & 4861.0  & 0.350 \\
H$\alpha$  & $3\rightarrow2$ & 6563.0  & 1.000 \\
Pa$\gamma$ & $6\rightarrow3$ & 10938.0 & 0.031 \\
Pa$\beta$  & $5\rightarrow3$ & 12822.0 & 0.057 \\
Pa$\alpha$ & $4\rightarrow3$ & 18756.0 & 0.119 \\
\hline
\end{tabular}

\vspace{0.5em}
\begin{minipage}{0.92\columnwidth}
\footnotesize
\textit{Note.} The listed wavelengths reproduce the rest-frame line centers adopted by the extraction pipeline. Hydrogen emissivity ratios relative to H$\alpha$ are derived from \citet{Osterbrock2006} under Case B recombination with $T_{\mathrm{e}}=10{,}000$ K and $n_{\mathrm{e}}=10^{4}\,\mathrm{cm}^{-3}$.
\end{minipage}
\end{table}

We remove spectral elements with non-finite wavelength, flux, or uncertainty values before attempting line measurements. We retain each spectrum on its native wavelength sampling. For each targeted transition, we define a local fitting window centered on its expected rest-frame laboratory wavelength. The window has a minimum half-width of 50~\AA\ and is expanded when necessary according to the local wavelength sampling. A preliminary width estimate of $\mathrm{FWHM}_{\mathrm{guess}}=5\Delta\lambda_{\mathrm{pix}}$, where $\Delta\lambda_{\mathrm{pix}}$ is the local spectral pixel scale in \AA\ per pixel, is used to determine the size of the fitting window. This preliminary estimate does not set the initial width of the Gaussian emission-line fit.

\begin{figure*}[ht!]
    \centering
    \includegraphics[width=\textwidth]{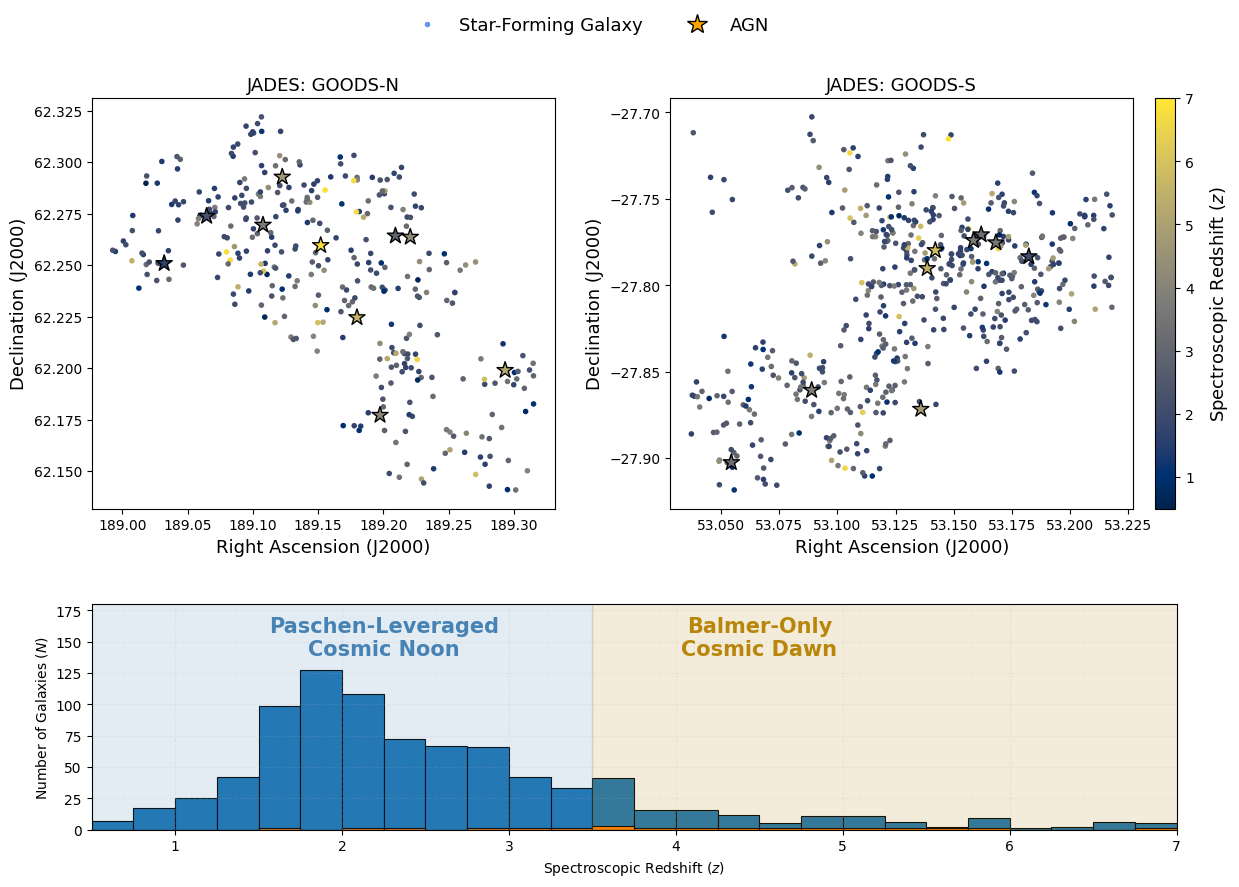}
    \caption{Top: Spatial distribution of the primary galaxy sample across the JADES GOODS-North and GOODS-South fields following the hierarchical AGN veto pipeline. Colored markers indicate the primary multi-line sample, colored by their spectroscopic redshift ($0.5 \le z \le 7$), with shapes denoting the final classification: retained star-forming galaxies (circles) and unambiguous hard AGN (stars). Bottom: Spectroscopic redshift distribution of the classified sample in bins of $\Delta z = 0.25$. Background shading explicitly delineates our two distinct analytical regimes: the Paschen-leveraged Cosmic Noon window ($0.5 \le z < 3.5$, blue) and the Balmer-only Cosmic Dawn window ($3.5 \le z \le 7$, tan). These classifications establish our pristine star-forming sample prior to the multi-line attenuation-solver convergence cuts described in Section~\ref{sec:analysis}.}
    \label{fig:scored_coverage}
\end{figure*}

To estimate the local continuum, we exclude a central region around the expected line position and fit the remaining sideband pixels with a linear function using a least-squares fit, which reduces the influence of  noise spikes. 
We subtract the resulting continuum before fitting the emission feature.

We estimate the local noise from the continuum-subtracted sidebands using the larger of a median-absolute-deviation estimate and the median spectral uncertainty, subject to a minimum noise floor. We then perform a matched-filter search over trial Gaussian dispersions of $\sigma=(0.6,\,0.9,\,1.3,\,1.9,\,2.7,\,3.8,\,5.4)\Delta\lambda_{\mathrm{pix}}$. The trial width that produces the highest matched-filter signal-to-noise ratio is adopted as the initial Gaussian width, with the corresponding matched-filter amplitude providing the initial estimate of the line amplitude.

We model each emission feature with a single positive Gaussian profile using bounded least squares with a soft-$L_1$ loss \citep[e.g.,][]{Cameron2023, Sanders2024}. The soft-$L_1$ loss behaves similarly to ordinary least squares for small residuals while reducing the influence of large residuals, thereby limiting the leverage of isolated outlying spectral pixels on the fitted profile.  We calculate integrated signal-to-noise ratios and retain lines with an integrated signal-to-noise ratio of $\mathrm{S/N}\geq3.5$. To prevent unrelated features from being identified as the targeted transition, we require the fitted centroid to remain within the larger of six spectral pixels or one-half of the fitted line FWHM from the expected rest-frame laboratory wavelength, such that $|\lambda_{\mathrm{fit}}-\lambda_{\mathrm{lab}}|\leq\max(6\Delta\lambda_{\mathrm{pix}},\,0.5\,\mathrm{FWHM})$.

\section{Sample Selection}
\label{sec:sample_selection}

We restrict the parent catalog of 3,156 unique galaxies to sources with spectroscopic redshifts in the range $0.5\leq z\leq7$, yielding 2,972 galaxies. This interval encompasses two observational regimes for the multi-line attenuation analysis. At $0.5\leq z\leq3.5$, the NIRSpec wavelength coverage can include both Balmer and Paschen transitions. At $3.5<z\leq7$, the measurements are primarily limited to Balmer transitions, allowing the method to be extended toward to high redshift but with reduced wavelength leverage.

Following the redshift selection, we require measurements of H$\alpha$, H$\beta$, and at least one additional hydrogen transition; this reduces our sample to 2,097 galaxies. Of these, 1,356 have Balmer Decrements above 2.86.  We then apply transition-specific requirements when selecting the third line: $\mathrm{S/N}\geq3.5$ for Pa$\beta$ and Pa$\alpha$, $\mathrm{S/N}\geq6$ for Pa$\gamma$, and $\mathrm{S/N}\geq8$ for H$\gamma$ and H$\delta$. These thresholds are intended to prevent a marginal third-line measurement from disproportionately influencing the recovered attenuation parameters. Pa$\beta$ and Pa$\alpha$ provide substantial wavelength leverage relative to the Balmer lines and are therefore retained at the standard detection threshold. Pa$\gamma$ is intrinsically weaker and is required to have a more secure detection. We use H$\gamma$ and H$\delta$ when no Paschen transition is available; because their shorter wavelength separation from H$\alpha$ and H$\beta$ provides less independent leverage on the attenuation slope and produces stronger covariance between $\tau_V$ and $n$, we adopt the most conservative signal-to-noise requirement for these transitions. These requirements result in 867 galaxies, which we we then cull for AGN.

\subsection{AGN Identification}
\label{sec:agn_veto}

Accurate dust attenuation modeling requires minimizing contamination from AGN. We remove AGN based on "hard" indicators. Optical strong-line ratios become highly degenerate between stellar and AGN ionizing sources at high ionization parameters and low metallicities \citep{cleri2025}. We therefore exclude a galaxy if it satisfies a single hard AGN criterion, or if it simultaneously satisfies at least two soft criteria.

\textbf{X-ray Luminosity (Hard Indicator).} We cross-match the optical coordinates of the galaxies with the deepest available X-ray catalogs, prioritizing the Chandra Deep Field South 7Ms and North 2Ms catalogs \citep{Luo2017, Xue2016}, supplemented by the Chandra Source Catalog 2.1 \citep[CSC 2.1;][]{Evans2010} for coverage outside these footprints. We classify sources with $L_X>10^{42}\,\mathrm{erg\,s^{-1}}$ as AGN. 

\textbf{Literature BLAGN (Hard Indicator).} We exclude any source previously confirmed as a broad-line AGN in JWST spectroscopic catalogs. Specifically, we cross-match our sample against the comprehensive JADES broad-line AGN census compiled by \citet{Juodzbalis2025}.




\section{Multi-Line Hydrogen Recombination Method}
\label{sec:analysis}

\subsection{Multi-line Dust Attenuation Framework}

We quantify dust attenuation by comparing the observed flux ratios of hydrogen recombination lines with their intrinsic ratios predicted under Case B recombination. We adopt an electron temperature of $T_{\mathrm e}=10{,}000\,\mathrm{K}$ and an electron density of $n_{\mathrm e}=10^{4}\,\mathrm{cm^{-3}}$ \citep{Osterbrock2006}. 

In the presence of dust, the observed flux of an emission line at wavelength $\lambda$ is related to its intrinsic flux through the effective optical depth \citep[e.g.,][]{Charlot2000}:

\begin{equation}
F_{\mathrm{obs}}(\lambda)=F_{\mathrm{int}}(\lambda)e^{-\tau(\lambda)}
\label{eqn:flux_dust_cor}
\end{equation}

\noindent where $F_{\mathrm{int}}(\lambda)$ is the intrinsic line flux and $\tau(\lambda)$ is the effective optical depth along the line of sight. Following \citet{Prescott2022}, we parameterize the wavelength dependence of the optical depth as a power law,

\begin{equation}
\tau(\lambda)=\tau_V\left(\frac{\lambda}{5500\,\text{\AA}}\right)^{-n}
\label{eqn:attenuation_law}
\end{equation}

\noindent where $\tau_V$, the optical depth at 5500A, sets the normalization of the attenuation curve and $n$ determines its wavelength-dependent slope. Larger values of $\tau_V$ correspond to greater overall attenuation, while larger values of $n$ produce a more steeply wavelength-dependent attenuation curve.

\begin{figure*}[t]
\centering
\includegraphics[width=0.9\textwidth]{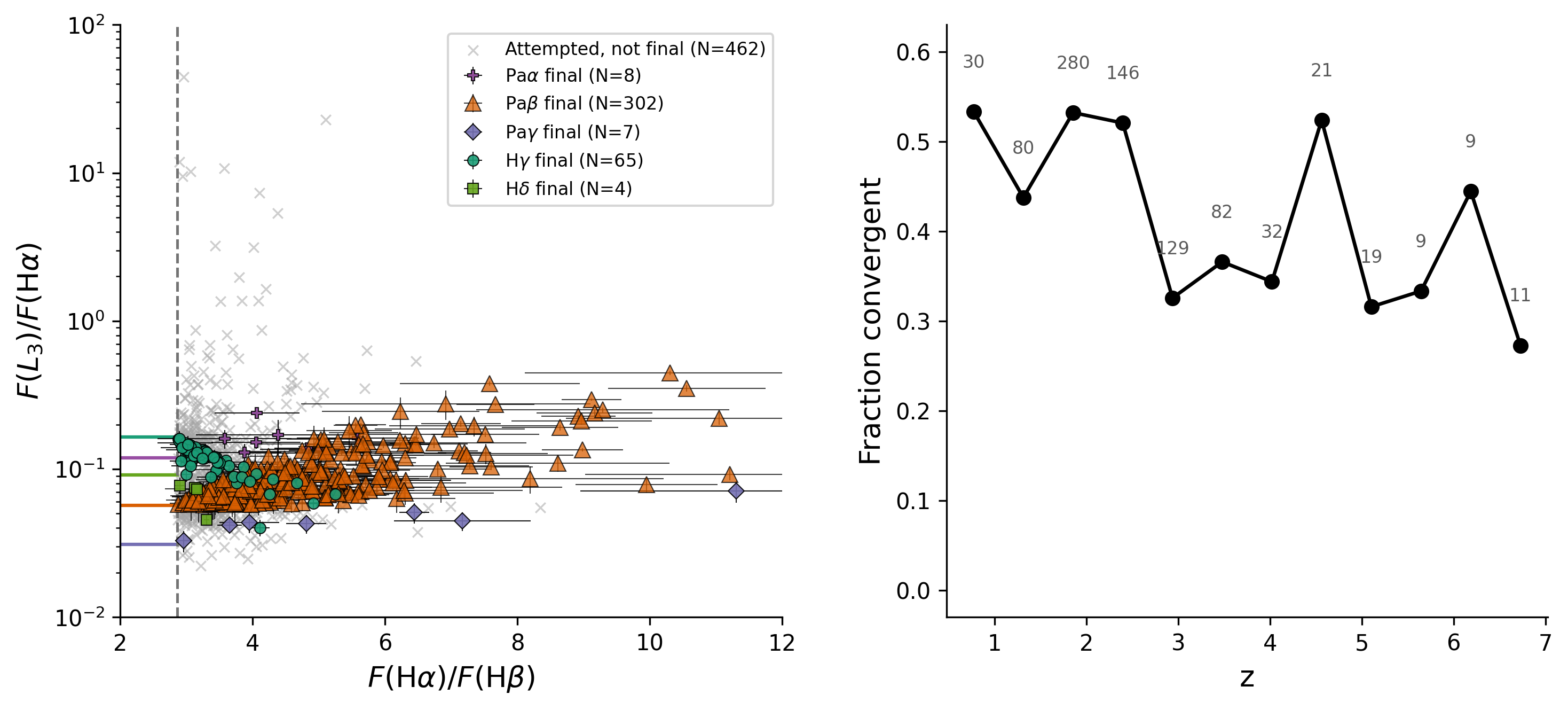}
\caption{
Left: Observed Balmer decrement, $F(\mathrm{H}\alpha)/F(\mathrm{H}\beta)$, compared with $F(L_3)/F(\mathrm{H}\alpha)$ for the 848 galaxies for which an attenuation solution is attempted, where $L_3$ denotes the third hydrogen recombination line used in the fit. Gray points show attempted solutions that do not converge or fail the adopted realism requirements, while colored symbols show accepted solutions according to the selected third line. The dashed vertical line marks the intrinsic Balmer decrement. The solid lines to the left of the Balmer decrement are the intrinsic ratio for the $F(L_3)/F(\mathrm{H}\alpha)$. Right: Fraction of attempted solutions that converge and satisfy the realism requirements as a function of redshift. Right: The number of attempted solutions in each redshift bin is shown above the corresponding point. The declining and increasingly variable convergence fraction at higher redshift is coupled to changes in line availability, wavelength leverage, sample size. However, galaxies will also fail if our assumed dust attenuation shape does not model the galaxies dust profile, therefore can indicate a change in shape of the dust attenuation with redshift.
}
\label{fig:tauv_n_conv_evo}
\end{figure*}

The observed ratio between two emission lines at rest-frame wavelengths $\lambda_i$ and $\lambda_j$ is

\begin{equation}
\frac{F_{\mathrm{obs}}(\lambda_i)}{F_{\mathrm{obs}}(\lambda_j)}
=
\frac{F_{\mathrm{int}}(\lambda_i)}{F_{\mathrm{int}}(\lambda_j)}
\exp\left\{-\left[\tau(\lambda_i)-\tau(\lambda_j)\right]\right\}
\label{eqn:flux_dust_ratio}
\end{equation}

\noindent where the subscripts $i$ and $j$ denote any two hydrogen recombination lines.  In our implementation, the two ratios are the Balmer decrement, $F(\mathrm{H}\alpha)/F(\mathrm{H}\beta)$, and the ratio of the selected third line to H$\alpha$, $F(L_3)/F(\mathrm{H}\alpha)$.

When multiple eligible third lines are available, we apply the fixed priority Pa$\beta$, Pa$\gamma$, Pa$\alpha$, H$\gamma$, and H$\delta$. This ordering reflects the combination of distance from H$\alpha$ and spectral availability. Only one third line enters the attenuation solution, additional hydrogen transitions are reserved for testing the resulting attenuation correction.

Substituting the power-law attenuation model from Equation~\ref{eqn:attenuation_law} into Equation~\ref{eqn:flux_dust_ratio} 
gives

\begin{equation}
\tau_{V,\mathrm{B}}(n)
=
\frac{\ln\left(R_{\mathrm{B}}/R_{\mathrm{B,int}}\right)}
{\left(\lambda_{\mathrm{H}\beta}/5500\,\text{\AA}\right)^{-n}
-
\left(\lambda_{\mathrm{H}\alpha}/5500\,\text{\AA}\right)^{-n}}
\label{eqn:tau_balmer}
\end{equation}

\noindent Where  $R_{\mathrm{B}}=F_{\mathrm{obs}}(\mathrm{H}\alpha)/F_{\mathrm{obs}}(\mathrm{H}\beta)$ and $R_{\mathrm{B,int}}=F_{\mathrm{int}}(\mathrm{H}\alpha)/F_{\mathrm{int}}(\mathrm{H}\beta)$. Similarly, defining $R_3=F_{\mathrm{obs}}(L_3)/F_{\mathrm{obs}}(\mathrm{H}\alpha)$ and $R_{3,\mathrm{int}}=F_{\mathrm{int}}(L_3)/F_{\mathrm{int}}(\mathrm{H}\alpha)$,the third-line ratio gives

\begin{equation}
\tau_{V,3}(n)
=
\frac{\ln\left(R_3/R_{3,\mathrm{int}}\right)}
{\left(\lambda_{\mathrm{H}\alpha}/5500\,\text{\AA}\right)^{-n}
-
\left(\lambda_{L_3}/5500\,\text{\AA}\right)^{-n}}
\label{eqn:tau_third}
\end{equation}

A physically consistent solution requires the two ratios to yield the same normalization. We define $f(n)=\tau_{V,\mathrm{B}}(n)-\tau_{V,3}(n)$ and solve for the value of $n$ at which $f(n)=0$.

We solve this nonlinear equation with a multi-start Newton--Raphson procedure. At iteration $k$, the attenuation slope is updated according to

\begin{equation}
n_{k+1}=n_k-\frac{f(n_k)}{f^{\prime}(n_k)}
\label{eqn:newton_update}
\end{equation}

\noindent where the derivative $f^{\prime}(n)$ is evaluated numerically. To mitigate systematic error or bias imposed by choice of initial conditions, 
we sequentially attempt the solution using starting values of $n=0.5$, $1.0$, $1.5$, $2.0$, $3.0$, $10.0$, and $20.0$ retaining the first solution that converges. Each starting value is allowed a maximum of 200 Newton--Raphson iterations. During the iteration, $n$ is restricted to the numerical interval $0\leq n\leq30$, and convergence is reached when $|n_{k+1}-n_k|<10^{-9}$. If a trial does not satisfy this criterion within 200 iterations, the solver proceeds to the next initial value. Once $n$ has converged, we calculate $\tau_V$ from Equation~\ref{eqn:tau_balmer}. 

We retain only solutions satisfying the bounds of $0\leq n\leq30$ and $0\leq\tau_V\leq30$. Solutions outside these intervals are mathematically permitted but are excluded from the high-confidence sample because they fall outside the physically plausible parameter space.

To assess how measurement uncertainties affect the recovered parameters, we resample Ha, Hb, and the third line fluxes using a Monte Carlo simulation 5000 times. The 16th and 84th percentiles of the resulting distributions define the uncertainties.

\begin{figure*}[t]
\centering
\includegraphics[width=\textwidth]{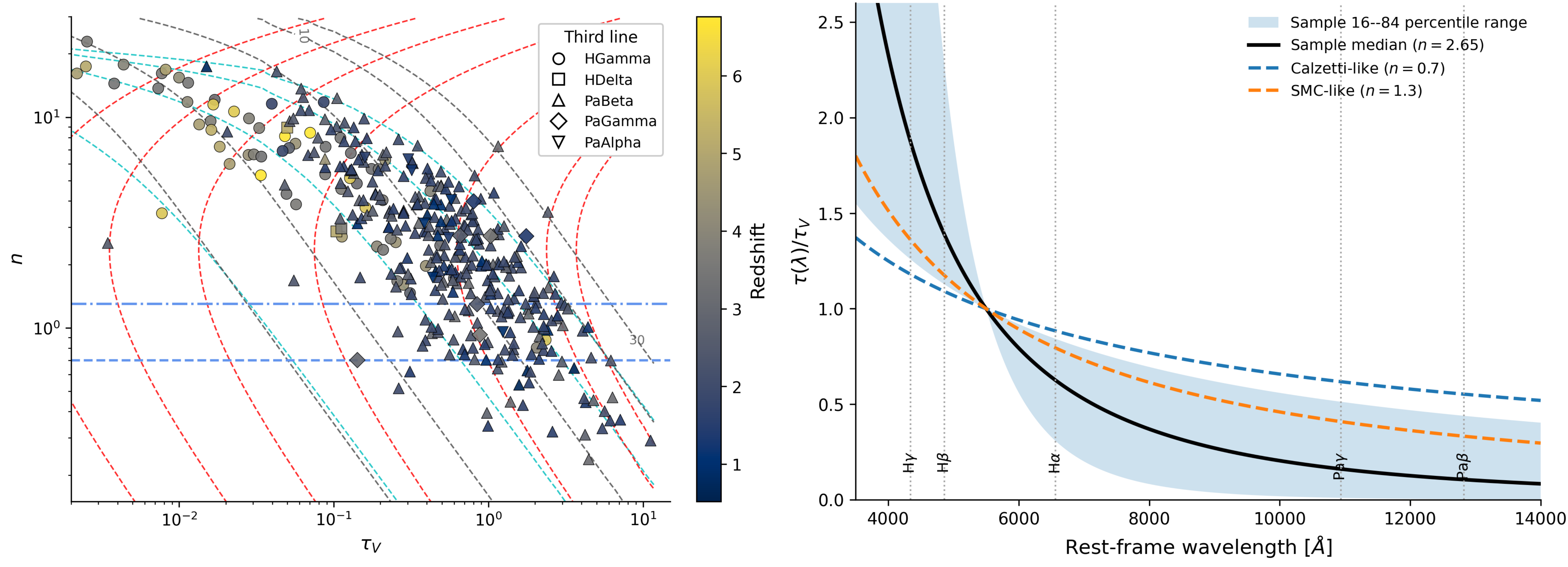}
\caption{
Left: Recovered attenuation normalization $\tau_V$ versus attenuation slope $n$ for the final galaxy sample. Marker shape denotes the third hydrogen recombination line used in the attenuation fit, while point color indicates redshift. Grey, red, and ligtht blue lines represent constant ratios of $F(\mathrm{H}\alpha)/F(\mathrm{H}\beta)$, $F(\mathrm{{Pa}}\beta)/F(\mathrm{H}\alpha)$, and $F(\mathrm{{H}}\gamma)/F(\mathrm{H}\alpha)$ respectively. 
Horizontal comparison lines indicate representative Calzetti-like ($n=0.7$) and SMC-like ($n=1.3$) slopes. Individual uncertainty regions are omitted for clarity; representative Monte Carlo distributions illustrating the covariance between $\tau_V$ and $n$ are presented in the Appendix. Right: Illustrative power-law attenuation curves corresponding to the recovered slope distribution. The solid black curve shows the sample median slope, $n=2.65$, and the shaded region represents the 16th--84th percentile range. The dashed comparison curves show the Calzetti-like and SMC-like slopes. 
}
\label{fig:tauv_n_conv}
\end{figure*}



Of the 848 galaxies, 386 converge and satisfy the realism requirements. Of the remainder, 345 fail to converge and 117 reach the upper boundary at $n=30$. 

\begin{table}[t]
\centering
\footnotesize
\caption{Sample Accounting}
\label{tab:sample_accounting}
\begin{tabular}{@{}lr@{}}
\hline
Selection or Analysis Stage & Number \\
\hline
Parent spectroscopic catalog & 3156 \\
Within $0.5\leq z\leq7$ & 2972 \\
Initial multi-line candidate sample & 867 \\
\hline
\multicolumn{2}{c}{\textbf{AGN Classification}} \\
\hline
Star-forming candidates & 848 \\
AGN & 19 \\
\hline
\multicolumn{2}{c}{\textbf{Attenuation-Solver Selection}} \\
\hline
Attenuation solutions attempted & 848 \\
\shortstack[l]{Convergent solutions passing\\realism cuts} & 386 \\
\hline
\end{tabular}
\end{table}

\subsection{Multi-line Dust Attenuation Analysis}

The left panel of Figure~\ref{fig:tauv_n_conv_evo} shows the observed line-ratios of all 848 galaxies. The 462 galaxies that do not produce an accepted solution are shown in gray, while the 386 galaxies that converge and pass the realism requirements are distinguished by the third hydrogen recombination line used in the fit. The accepted solutions include 302 (49.2\%) galaxies using Pa$\beta$, 7 (24.1\%) using Pa$\gamma$, 8 (42.1\%) using Pa$\alpha$, 65 (39.4\%) using H$\gamma$, and 4 (19.0\%) using H$\delta$. The percentages indicate the fraction of the galaxies with that third hydrogen line that converge.

The distribution in the left panel of Figure~\ref{fig:tauv_n_conv_evo} illustrates the importance of wavelength leverage. Solutions using Paschen transitions, particularly Pa$\beta$, occupy a broader range in $F(L_3)/F(\mathrm{H}\alpha)$ than those using H$\gamma$ or H$\delta$. The fact H$\gamma$ is tightly correlated with the Balmer Decrement at high optical depths, shows why the use of another Balmer line is not able to produce a larger sample of converging slope and normalizations. We also find the expected optical depth at 5500\AA reduce as we reach higher redshifts, as redshift increases we find that dust attenuation slopes also increase.

Among the available third-line choices, Pa$\beta$ provides the most useful combination of wavelength leverage, spectral availability, and SNR, resulting in a larger and more reliable sample size. The representative Monte Carlo distributions in the Appendix also show that Pa$\beta$ produces a more localized solution region than the Balmer-only combinations.  
Balmer-only solutions can extend the analysis to higher redshift, but their recovered slopes should be interpreted more cautiously because the shorter wavelength baseline produces stronger covariance between attenuation normalization and slope.

We next examine the convergence fraction as a function of redshift in the right panel of Figure~\ref{fig:tauv_n_conv_evo}. At $z\lesssim2.5$, approximately 50\% of the attempted fits in each bin produce accepted solutions. At higher redshifts, the convergence fraction becomes more variable and generally declines. 
This primarily will be due to Paschen transitions move outside the NIRSpec wavelength range above $z\sim3.5$, and H$\delta$ or H$\gamma$ are less likely to converge on solutions. 

The best-fit parameters for the final sample are shown in the left panel of Figure~\ref{fig:tauv_n_conv}. The solutions span nearly the full allowed range in attenuation slope and exhibit a broad inverse relationship between $n$ and $\tau_V$. For example, near $\tau_V\approx0.1$, the accepted slopes extend approximately from $n\approx5-20$, whereas galaxies with $\tau_V\gtrsim4$ are concentrated primarily at $n\lesssim1$. Our results occupy a similar region to \cite{Prescott2022},  where their four galaxies at $z \approx 0.3–2$, spands $0<n<1.6$ and $1<\tau_V<6.8$. The 54 galaxies at $z>3.5$ have average recovered values of $n\approx6.48$ and $\tau_V\approx0.05$, compared with $n\approx2.2$ and $\tau_V\approx0.58$ at $z<3.5$. 

Only 17.1\% of the final galaxies have slopes within the commonly adopted interval $0.7\leq n\leq1.3$. The median recovered slope is $n=2.65$, which is steeper than the Calzetti-like value $n=0.7$ and slightly steeper than the SMC-like value $n=1.3$ shown in the right panel of Figure~\ref{fig:tauv_n_conv}. 
The shaded 16th--84th percentile region extends across both canonical prescriptions and to substantially steeper slopes. 

Our galaxies are selected through emission lines and the current analysis does not include stellar-mass measurements. We cannot determine which systems qualify as starbursts relative to galaxies of similar mass and redshift. We therefore cannot assume it is directly analogous to the local starburst population for which the Calzetti attenuation curve was calibrated.

The broad slope distribution is consistent with previous evidence that attenuation curves vary among galaxies and correlate with dust opacity \citep[e.g.,][]{Reddy2015,Salim_2020}. For example \cite{Reddy2015} found scatter in the UV slope at fixed Balmer optical depth, with a population-averaged attenuation curve 
which was similar to the SMC curve $\lambda \gtrsim$ 2500 Å but was Calzetti-like in the UV,
for star-forming galaxies at $1.4<z<2.6$.

\subsection{Limitations and Systematic Uncertainties}

While the multi-line attenuation framework constrains both the normalization and slope of the attenuation curve, several systematic uncertainties remain. The intrinsic hydrogen-line ratios adopted in this analysis assume Case B with $T_{\mathrm e}=10{,}000\,\mathrm{K}$ and $n_{\mathrm e}=10^{4}\,\mathrm{cm^{-3}}$. Deviations from this assumed electron temperature and density may be particularly relevant at high redshift, where lower metallicities and harder ionizing radiation fields can produce higher electron temperatures. A mismatch between the adopted and true intrinsic line ratios would shift the inferred values of $\tau_V$ and $n$. The current analysis does not include independent electron-temperature or density measurements, so we cannot correct this effect. Future applications could reduce this uncertainty by adopting intrinsic recombination ratios derived from measured nebular conditions when temperature- or density-sensitive diagnostics are available.

The recovered attenuation parameters also represent effective, line-of-sight averages rather than a unique measurement of the underlying dust geometry. Real galaxies can contain clumpy dust distributions, differential obscuration among star-forming regions, and mixtures of emitting gas and absorbing material that are not fully described by a single foreground-screen power-law \cite[e.g.][]{Forrey2025}. \cite{Nelson_2016} shows such spatially resolved attenuation analysis, mapping H$\alpha$ and H$\beta$ strengths in 609 galaxies at $z\sim 1.4$. This team found radially varied Balmer decrement, suggesting deeper geometric attenuation influences that cannot always be characterized by a single dust curve. 

One indication that the adopted model is inadequate is the failure of a galaxy's three measured line ratios to produce an accepted solution within the adopted parameter bounds. However, neither non-convergence nor residual line-to-line inconsistency uniquely identifies complex geometry as the cause, because measurement errors, departures from the adopted Case B ratios, and spectral contamination can produce similar behavior. Distinguishing among these possibilities would require additional constraints, such as spatially resolved spectroscopy, or independent continuum-based attenuation measurements. 

This is not a homogeneously selected sample, so we can't draw general conclusions about the dust attenuation parameters for all galaxies. Rather, this study illustrates the variety of parameter values and indicates that there is no universal dust law that can be adopted.

\section{Resulting Star Formation Rates} 
\label{sec:SFR}

\subsection{Hydrogen-Line SFR Conversions and Uncorrected Baseline}
\begin{figure*}[t]
    \centering
    \includegraphics[width=\textwidth]{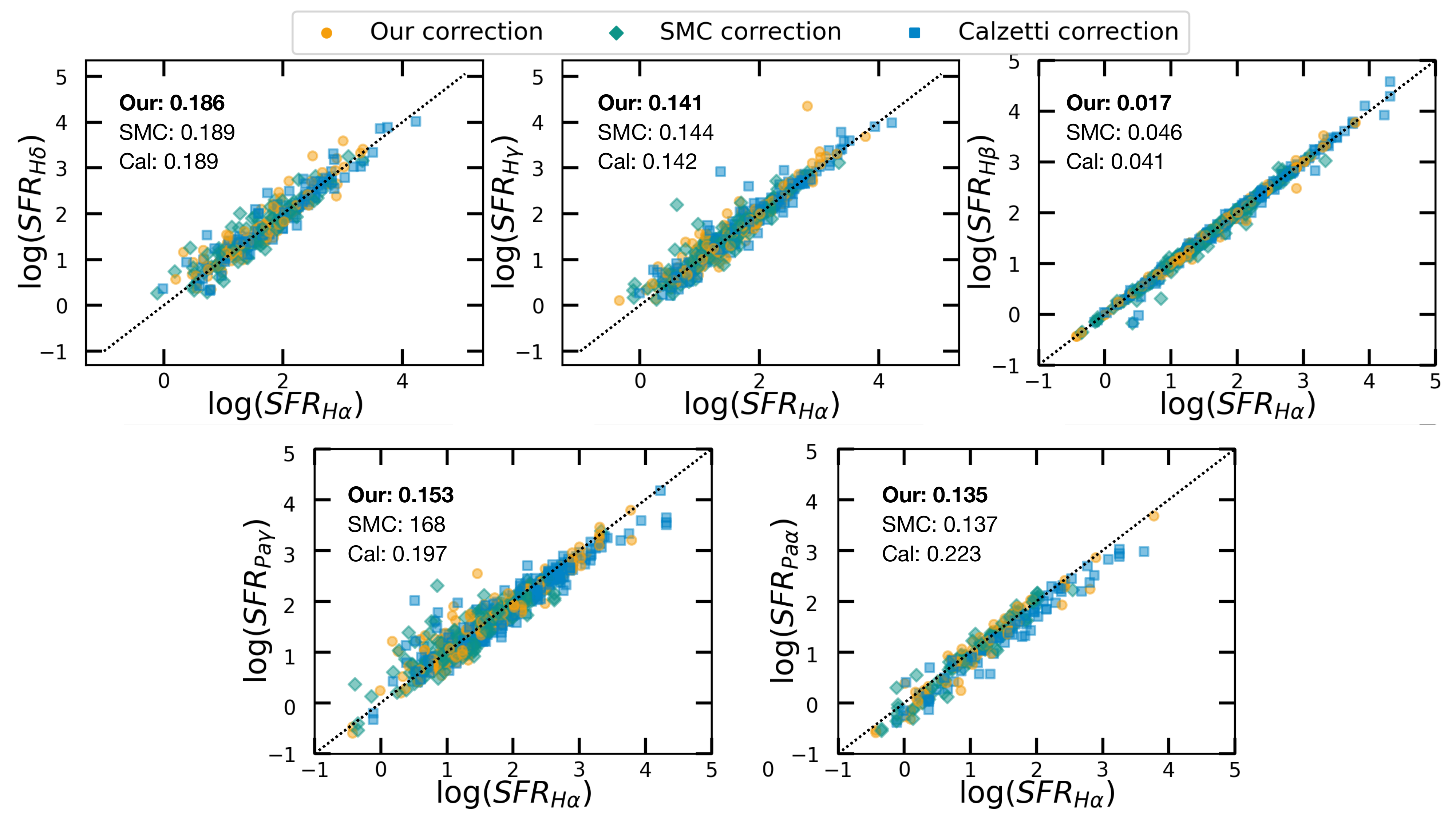}
    \caption{
    Consistency of SFRs inferred from different attenuation corrections, multi-line dust attenution (This work), SMC and Calzetti-like. The dotted line represents unity and the annotation represent the scatter of the method from unity. All methods rely on the Balmber decrement (upper right panel), however the other panels do not include SFR's from galaxies that used that third line to constrain the normalization and slope. This allows a less bias comparison. 
    }
    \label{fig:SFR_Ratios}
\end{figure*}

To characterize the recent star formation activity of our convergent galaxy sample, we derive SFRs from H$\alpha$, H$\beta$, and Pa$\beta$. Because the three transitions arise from the same ionized-gas emission but span a wide wavelength range, they are expected to yield consistent SFRs while experiencing substantially different levels of dust attenuation. Comparing their SFRs therefore provides a sensitive test of the adopted attenuation correction.

The H$\alpha$ luminosities are converted to SFRs following the calibration of \citet{Kennicutt2012}:

\begin{equation}
\text{SFR}_{\mathrm{H}\alpha}\,[M_\odot\,\mathrm{yr}^{-1}]
=
5.37 \times 10^{-42} \,
L_{\mathrm{H}\alpha}\,[\mathrm{erg\,s^{-1}}]
\label{Eqn:Kennicutt_SFR}
\end{equation}

We derive corresponding SFR conversions for the other hydrogen recombination lines using the intrinsic Case B emissivity ratios adopted throughout this work. For a transition H$\lambda$, the intrinsic luminosity ratio is

\begin{equation}
\frac{L_{\mathrm{int}}(\mathrm{H}\alpha)}
{L_{\mathrm{int}}(\mathrm{H}\lambda)}
=
\frac{j_{\mathrm{H}\alpha}}
{j_{\mathrm{H}\lambda}},
\label{Eqn:intrinsic_Lum_Ratio}
\end{equation}

\noindent where $j_{\mathrm{H}\lambda}$ is the Case B emissivity of the target transition for $T_{\mathrm e}=10{,}000\,\mathrm{K}$ and $n_{\mathrm e}=10^{4}\,\mathrm{cm^{-3}}$ \citep{Osterbrock2006}. Combining Equation~\ref{Eqn:intrinsic_Lum_Ratio} with the H$\alpha$ calibration gives

\begin{equation}
\mathrm{SFR}_{\mathrm{H}\lambda}\,[M_\odot\,\mathrm{yr}^{-1}]
=
5.37\times10^{-42}
\left(
\frac{j_{\mathrm{H}\alpha}}
{j_{\mathrm{H}\lambda}}
\right)
L_{\mathrm{H}\lambda}\,[\mathrm{erg\,s^{-1}}].
\label{Eqn:any_SFR}
\end{equation}

The conversion factors for H$\beta$, Pa$\beta$, and the remaining hydrogen transitions follow directly from the emissivity ratios listed in Table~\ref{tab:emission_lines}. Using a common H$\alpha$ calibration ensures that, in the absence of measurement error and departures from  Case B conditions, the intrinsic luminosities of all hydrogen lines yield the same SFR.

Because observed luminosities are attenuated by dust, using Equation \ref{eqn:attenuation_law}, the corrected SFR estimate is:

\begin{equation}
\mathrm{SFR}_{H\lambda}
= 5.37\times10^{-42}
\left(\frac{j_{H\alpha}}{j_{H\lambda}}\right)
L_{\mathrm{obs},H\lambda}
e^{\tau_{\lambda,H\lambda}},
\label{Eqn:intrinsic_SFR_Ratio}
\end{equation}


\noindent where $\tau_{\nu,H\lambda}$ is the attenuation. 


If the attenuation correction accurately recovers the intrinsic line luminosities, the independently inferred SFRs approach equality. Therefore, we can use additional hydrogen lines that were not used in the fit to provide a second diagnostic of model inadequacy. 
Figure~\ref{fig:SFR_Ratios} compares the SFRs from different lines.

We compare the SFRs from our attenuation models, the Calzetti curve, and the SMC curve.
The median scatter between all SFR ratios, using our galaxy-specific attenuation models,  is $0.07\,\mathrm{dex}$, whereas the median scatter between all SFR ratios for SMC dust corrections is  $0.09\,\mathrm{dex}$ and Calzetti-like corrections is $0.11\,\mathrm{dex}$. 
When considering all wavelengths, Calzetti-like corrections, $\sim5.5$\% of galaxies are off by a factor of 2. For SMC corrections, $\sim6.2$\% of galaxies are off by a factor of 2. For our corrections, $\sim7.5$\% of galaxies are off by a factor of 2. 



The direction and magnitude of this discrepancy depend on the assumed slope. A shallower Calzetti-like curve requires a larger $\tau_V$ to reproduce a given Balmer decrement and therefore predicts more attenuation at $\lambda\geq$Pa$\gamma$, overinflating the SFR. The steeper SMC-like prescription reduces this effect but still produces residuals for galaxies whose attenuation slopes differ from the fixed value. These results demonstrate that neither fixed prescription represents the full range of attenuation behavior in the sample. However, if it is not possible to individually measure attenuation curves, the SMC model matches most closely to our results, and we recommend its usage at $z>0.5$

The galaxy-specific curves provide more consistent attenuation corrections across the broader set of available hydrogen transitions than either uncorrected luminosities or fixed Calzetti-like and SMC-like prescriptions. The diversity in the recovered attenuation slopes has measurable consequences for nebular SFR estimates.

\section{Summary and Conclusion}
\label{sec:summary}

We use JWST/NIRSpec spectroscopy from JADES to investigate effective dust attenuation curves in 386 emission-line-selected star-forming galaxies at $0.5\leq z\leq7$. The sample includes a Paschen-leveraged Cosmic Noon regime at $1\leq z\leq3.5$ and a Balmer-only Cosmic Dawn regime at $3.5<z\leq7$. For each galaxy, we combine H$\alpha$, H$\beta$, and one additional hydrogen recombination line to constrain the normalization $\tau_V$ and slope $n$ of a power-law attenuation model. Unlike conventional Balmer-decrement corrections, this approach assumes a common functional form but allows the attenuation slope to vary from galaxy to galaxy rather than fixing it to a canonical empirical value.

The principal results are as follows:

\begin{enumerate}

\item The recovered attenuation curves exhibit substantial galaxy-to-galaxy diversity. The median attenuation slope is $n=2.65$, and only 17.1\% of the final sample lies within the commonly adopted interval $0.7\leq n\leq1.3$. The recovered distribution includes curves that are both shallower and substantially steeper than the Calzetti-like and SMC-like prescriptions, demonstrating that no single fixed slope represents the full sample.

\item Paschen transitions provide considerably greater leverage on attenuation-curve shape than Balmer-only combinations. Their longer wavelength separation from H$\alpha$ and H$\beta$ helps distinguish attenuation normalization from slope, whereas solutions based on H$\gamma$ or H$\delta$ remain more strongly covariant. 

\item Fixed Calzetti-like and SMC-like do not consistently reconcile the SFRs inferred from optical and near-infrared hydrogen lines. With a higher percent of galaxies with disagreements over a factor of 2, showing that an attenuation curve anchored only to H$\alpha$/H$\beta$ can produce biased corrections.

\item The galaxy-specific corrections bring the three fitted hydrogen lines into mathematical agreement and reduce the line-to-line scatter among additional hydrogen transitions that were not used to determine $\tau_V$ and $n$. The comparison indicates that the recovered curves predict attenuation across a broader wavelength range more consistently than either uncorrected luminosities or fixed-slope prescriptions.

\item The higher-redshift subset has, on average, steeper recovered slopes and lower optical depths than the lower-redshift sample. However, this apparent trend is based on a small Balmer-only subset and is coupled to changing line availability, wavelength leverage, and measurement quality. It should therefore be interpreted as tentative evidence for evolution rather than as a direct measurement of changes in metallicity, dust-grain populations, or star--dust geometry.

\end{enumerate}

Diversity in effective attenuation-curve shape has measurable consequences for nebular SFR estimates. Multi-line spectroscopy extending from the Balmer to the Paschen series allows the attenuation slope to be constrained directly when suitable lines are available and provides more consistent galaxy-specific corrections across the hydrogen recombination spectrum. Expanding this approach to larger high-redshift samples with stronger long-wavelength line coverage will be necessary to determine whether the apparent redshift dependence reflects genuine evolution in dust properties.


\section{Acknowledgments}
We acknowledge the contributions of everyone involved in the JWST mission.
O.Z., J.R.H., B.E.B., and A.K. acknowledge support from JWST-GO-03794.
Some of the data products presented herein were retrieved from the DAWN JWST Archive (DJA). DJA is an initiative of the Cosmic Dawn Center (DAWN), which is funded by the Danish National Research Foundation under grant DNRF140.

\facilities{JWST(NIRSpec) \citep{2022A&A...661A..80J}}

\software{astropy \citep{2013A&A...558A..33A, 2018AJ....156..123A, 2022ApJ...935..167A},
          numpy \citep{harris2020array},
          scipy \citep{2020SciPy-NMeth},
          matplotlib \citep{Hunter:2007},
          }

\restartappendixnumbering
\appendix

\section{Additional Diagnostic and Uncertainty Figures}
\label{sec:appendix}

\begin{figure*}[h]
\centering
\includegraphics[width=\textwidth]{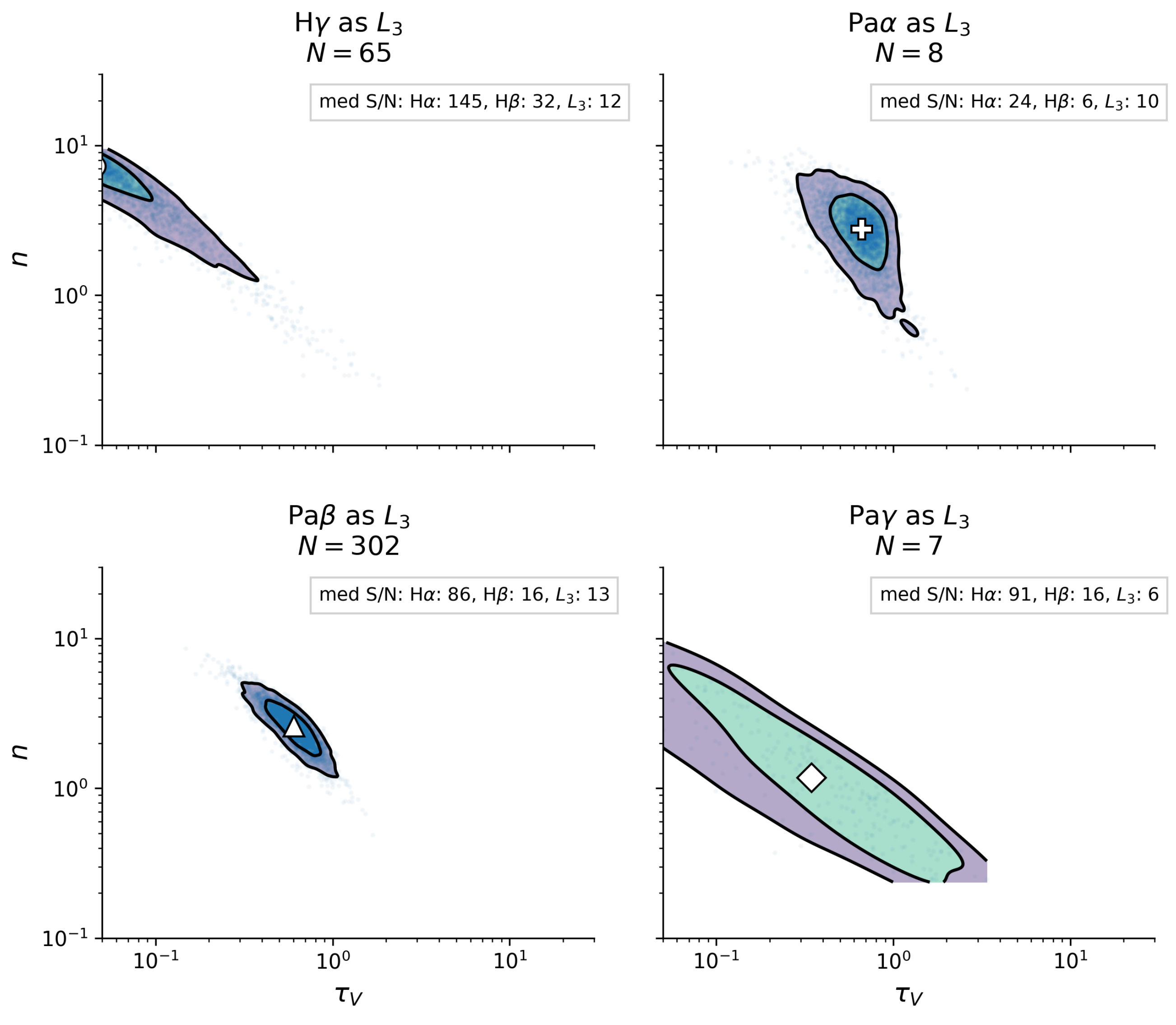}
\caption{Monte Carlo uncertainty distributions in attenuation-parameter space for fits using H$\gamma$, Pa$\alpha$, Pa$\beta$, and Pa$\gamma$ as the third hydrogen recombination line. 
Each distribution contains 5,000 successful solutions obtained by perturbing the three input line fluxes with Gaussian draws based on their measurement uncertainties and rerunning the attenuation solver. Filled contours show the highest-density regions of the recovered $(\tau_V,n)$ distributions, and the large symbol marks the solution obtained from the unperturbed representative fluxes. The elongated contours illustrate the covariance between attenuation normalization and slope. The Pa$\beta$ example produces the most localized representative distribution, while the other panels remain broader because of differences in wavelength leverage and signal-to-noise ratio.
}
\label{fig:appendix_mc_clouds}
\end{figure*}

This appendix shows the Monte Carlo uncertainty propagation described in Section~\ref{sec:analysis}. Figure~\ref{fig:appendix_mc_clouds} shows representative solution distributions in $(\tau_V,n)$ space for fits using H$\gamma$, Pa$\alpha$, Pa$\beta$, and Pa$\gamma$ as the third hydrogen recombination line.  The measured fluxes are perturbed repeatedly within their uncertainties before rerunning the attenuation solver.

The elongated distributions illustrate the covariance between attenuation normalization and slope. The Pa$\beta$ solution is comparatively localized, while the H$\gamma$, Pa$\alpha$, and Pa$\gamma$ examples span broader regions of parameter space. These differences reflect both the wavelength leverage of the selected transition and the signal-to-noise ratios of the measurements; they should therefore not be attributed to transition choice alone. The broad covariance is qualitatively consistent with \citet{Prescott2022}, who likewise found that observational uncertainties substantially weakened the constraints for most of their initially plausible three-line solutions.

\bibliography{ref}{}

@article{Kennicutt2012,
       author = {{Kennicutt}, R. C. and {Evans}, N. J.},
        title = "{Star Formation in the Milky Way and Nearby Galaxies}",
      journal = {\araa},
         year = 2012,
        month = sep,
       volume = {50},
        pages = {531-608},
          doi = {10.1146/annurev-astro-081811-125610}
}

@article{Calzetti2001,
       author = {{Calzetti}, Daniela},
        title = "{The Dust Content of Galaxies: A Review}",
      journal = {\pasp},
         year = 2001,
        month = may,
       volume = {113},
       number = {783},
        pages = {1449-1485},
          doi = {10.1086/324269},
       adsurl = {https://ui.adsabs.harvard.edu/abs/2001PASP..113.1449C}
}

@ARTICLE{Charlot2000,
       author = {{Charlot}, St{\'e}phane and {Fall}, S. Michael},
        title = "{A Simple Model for the Absorption of Starlight by Dust in Galaxies}",
      journal = {\apj},
         year = 2000,
        month = aug,
       volume = {539},
       number = {2},
        pages = {718-731},
          doi = {10.1086/309250},
archivePrefix = {arXiv},
       eprint = {astro-ph/0003128},
 primaryClass = {astro-ph},
       adsurl = {https://ui.adsabs.harvard.edu/abs/2000ApJ...539..718C}
}

@article{Gordon2003,
       author = {{Gordon}, Karl D. and {Clayton}, Geoffrey C. and {Misselt}, K.~A. and {Landolt}, A.~U. and {Wolff}, Michael J.},
        title = "{A Quantitative Comparison of the Small Magellanic Cloud, Large Magellanic Cloud, and Milky Way Ultraviolet to Near-Infrared Extinction Curves}",
      journal = {\apj},
         year = 2003,
        month = sep,
       volume = {594},
       number = {1},
        pages = {279-293},
          doi = {10.1086/376774},
archivePrefix = {arXiv},
       eprint = {astro-ph/0305257},
 primaryClass = {astro-ph},
       adsurl = {https://ui.adsabs.harvard.edu/abs/2003ApJ...594..279G}
}

@ARTICLE{Cameron2023,
       author = {{Cameron}, A.~J. and {Katz}, H. and {Rey}, J. and {Saxena}, A. and {Bunker}, A.~J. and {Carniani}, S. and {Charlot}, S. and {Curtis-Lake}, E. and {D'Eugenio}, F. and {Ferruit}, P. and {Jakobsen}, P. and {Maiolino}, R. and {Scholtz}, J. and {Tacchella}, S. and {Willott}, C.~J. and {Witstok}, J.},
        title = "{JADES: Probing interstellar dust at z \raisebox{-0.5ex}\textasciitilde 5.5-7.0 with empirical Case B recombination}",
      journal = {\aap},
         year = 2023,
        month = sep,
       volume = {677},
          eid = {A115},
        pages = {A115},
          doi = {10.1051/0004-6361/202346107}
}

@ARTICLE{Sanders2024,
       author = {{Sanders}, Ryan L. and {Shapley}, Alice E. and {Topping}, Michael W. and {Reddy}, Naveen A. and {Brammer}, Gabriel B.},
        title = "{A JWST/NIRSpec Reference Sample for the Physical Properties of Galaxies at z \raisebox{-0.5ex}\textasciitilde 1.5-9}",
      journal = {\apj},
         year = 2024,
        month = may,
       volume = {966},
       number = {1},
          eid = {73},
        pages = {73},
          doi = {10.3847/1538-4357/ad36c6}
}

@BOOK{Osterbrock2006,
       author = {{Osterbrock}, Donald E. and {Ferland}, Gary J.},
        title = "{Astrophysics of Gaseous Nebulae and Active Galactic Nuclei}",
         year = 2006,
    publisher = {University Science Books},
       adsurl = {https://ui.adsabs.harvard.edu/abs/2006agna.book.....O}
}

@article{Hickox2018,
       author = {{Hickox}, Ryan C. and {Alexander}, David M.},
        title = "{Obscured Active Galactic Nuclei}",
      journal = {\araa},
         year = 2018,
        month = aug,
       volume = {56},
        pages = {625-673},
          doi = {10.1146/annurev-astro-081817-051803},
archivePrefix = {arXiv},
       eprint = {1805.00465},
 primaryClass = {astro-ph.GA},
       adsurl = {https://ui.adsabs.harvard.edu/abs/2018ARA&A..56..625H}
}

@misc{Reddy2015,
      title={The MOSDEF Survey: Measurements of Balmer Decrements and the Dust Attenuation Curve at Redshifts z~1.4-2.6}, 
      author={Naveen A. Reddy and Mariska Kriek and Alice E. Shapley and William R. Freeman and Brian Siana and Alison L. Coil and Bahram Mobasher and Sedona H. Price and Ryan L. Sanders and Irene Shivaei},
      year={2015},
      eprint={1504.02782},
      archivePrefix={arXiv},
      primaryClass={astro-ph.GA},
      url={https://arxiv.org/abs/1504.02782}, 
}

@article{Nelson_2016,
   title={Spatially Resolved Dust Maps From Balmer Decrements In Galaxies At z ∼ 1.4},
   volume={817},
   ISSN={2041-8213},
   url={http://dx.doi.org/10.3847/2041-8205/817/1/L9},
   DOI={10.3847/2041-8205/817/1/l9},
   number={1},
   journal={The Astrophysical Journal Letters},
   publisher={American Astronomical Society},
   author={Nelson, Erica June and Dokkum, Pieter G. van and Momcheva, Ivelina G. and Brammer, Gabriel B. and Wuyts, Stijn and Franx, Marijn and Schreiber, Natascha M. Förster and Whitaker, Katherine E. and Skelton, Rosalind E.},
   year={2016},
   month=Jan, pages={L9} }

@article{Cleri2022,
       author = {{Cleri}, N. J. and {Trump}, J. R. and {Backhaus}, B. E. and {Momcheva}, I. and {Papovich}, C. and {Simons}, R. and {Weiner}, B. and {Estrada-Carpenter}, V. and {Finkelstein}, S. L. and {Giavalisco}, M. and {Ji}, Z. and {Jung}, I. and {Matharu}, J. and {Martinez}, F. and {Sturm}, M. R.},
        title = "{CLEAR: Paschen-$\beta$ Star Formation Rates and Dust Attenuation of Low Redshift Galaxies}",
      journal = {\apj},
         year = 2022,
        month = apr,
       volume = {929},
       number = {1},
          eid = {83},
        pages = {83},
          doi = {10.3847/1538-4357/ac5a4c},
archivePrefix = {arXiv},
       eprint = {2009.00617},
 primaryClass = {astro-ph.GA},
       adsurl = {https://ui.adsabs.harvard.edu/abs/2022ApJ...929...83C}
}

@article{DEugenio2024,
       author = {{D'Eugenio}, Francesco and others},
        title = "{JADES Data Release 3: NIRSpec/MSA spectroscopy for 4,000 galaxies in the GOODS fields}",
      journal = {arXiv e-prints},
         year = 2024,
        month = apr,
          eid = {arXiv:2404.06531},
archivePrefix = {arXiv},
       eprint = {2404.06531}
}

@article{deGraaff2024,
       author = {{de Graaff}, A. and others},
        title = "{JWST/NIRSpec spectroscopy of intermediate-mass quiescent galaxies at z ~ 3-4}",
      journal = {\mnras},
         year = 2024,
        month = aug,
       volume = {534},
        pages = {3552-3560},
          doi = {10.1093/mnras/stae2206}
}

@article{Heintz2023,
       author = {{Heintz}, Kasper E. and others},
        title = "{Extreme Dust Attenuation in an Epoch of Reionization Galaxy}",
      journal = {\apjl},
         year = 2023,
        month = jul,
       volume = {952},
       number = {2},
          eid = {L32},
        pages = {L32},
          doi = {10.3847/2041-8213/ace41d}
}

@article{Spergel2003,
       author = {{Spergel}, D.~N. and others},
        title = "{First-Year Wilkinson Microwave Anisotropy Probe (WMAP) Observations: Determination of Cosmological Parameters}",
      journal = {\apjs},
         year = 2003,
       volume = {148},
        pages = {175-194},
          doi = {10.1086/377226}
}

@article{Evans2010,
       author = {{Evans}, I.~N. and others},
        title = "{The Chandra Source Catalog}",
      journal = {\apjs},
         year = 2010,
       volume = {189},
        pages = {37-82},
          doi = {10.1088/0067-0049/189/1/37}
}

@ARTICLE{Forrey2025,
       author = {{Forrey}, Grace M. and {Simons}, Raymond C. and {Trump}, Jonathan R. and {Shen}, Lu and {Koekemoer}, Anton M. and {Bagley}, Micaela B. and {Finkelstein}, Steven L. and {Papovich}, Casey and {Pirzkal}, Nor},
        title = "{NGDEEP: A New Non-Parametric Measure of Local Star-Formation and Attenuation at Cosmic Noon}",
      journal = {arXiv e-prints},
         year = 2025,
        month = dec,
          eid = {arXiv:2512.11989},
        pages = {arXiv:2512.11989},
          doi = {10.48550/arXiv.2512.11989},
archivePrefix = {arXiv},
       eprint = {2512.11989},
 primaryClass = {astro-ph.GA},
       adsurl = {https://ui.adsabs.harvard.edu/abs/2025arXiv251211989F}
}

@misc{cleri2025,
      title={Optical Strong Line Ratios Cannot Distinguish Between Stellar Populations and Accreting Black Holes at High Ionization Parameters and Low Metallicities}, 
      author={Nikko J. Cleri and Grace M. Olivier and Bren E. Backhaus and Joel Leja and Casey Papovich and Jonathan R. Trump and Pablo Arrabal Haro and Veronique Buat and Denis Burgarella and Emilie Burnham and Antonello Calabro and Jonathan H. Cohn and Justin W. Cole and Kelcey Davis and Mark Dickinson and Steven L. Finkelstein and Ray Garner III and Michaela Hirschmann and Weida Hu and Taylor A. Hutchison and Dale D. Kocevski and Anton M. Koekemoer and Rebecca L. Larson and Zach J. Lewis and Michael V. Maseda and Lise-Marie Seille and Raymond C. Simons},
      year={2025},
      eprint={2506.21660},
      archivePrefix={arXiv},
      primaryClass={astro-ph.GA},
      url={https://arxiv.org/abs/2506.21660}, 
}

@article{Salim_2020,
   title={The Dust Attenuation Law in Galaxies},
   volume={58},
   ISSN={1545-4282},
   url={http://dx.doi.org/10.1146/annurev-astro-032620-021933},
   DOI={10.1146/annurev-astro-032620-021933},
   number={1},
   journal={Annual Review of Astronomy and Astrophysics},
   publisher={Annual Reviews},
   author={Salim, Samir and Narayanan, Desika},
   year={2020},
   month=Aug, pages={529–575} }

@article{Prescott2022,
       author = {{Prescott}, M. K. M. and {Finlator}, K. M. and {Cleri}, N. J. and {Trump}, J. R. and {Papovich}, C.},
        title = "{Using Multiple Emission Line Ratios to Constrain the Slope of the Dust Attenuation Law}",
      journal = {\apj},
         year = 2022,
        month = mar,
       volume = {928},
       number = {1},
          eid = {71},
        pages = {71},
          doi = {10.3847/1538-4357/ac5459},
archivePrefix = {arXiv},
       eprint = {2202.04107},
 primaryClass = {astro-ph.GA},
       adsurl = {https://ui.adsabs.harvard.edu/abs/2022ApJ...928...71P}
}

@article{Shapley2023,
       author = {{Shapley}, Alice E. and {Sanders}, Ryan L. and {Reddy}, Naveen A. and {Topping}, Michael W. and {Brammer}, Gabriel B.},
        title = "{JWST/NIRSpec Balmer-line Measurements of Star Formation and Dust Attenuation at z {\ensuremath{\sim}} 3–6}",
      journal = {\apj},
         year = 2023,
        month = sep,
       volume = {954},
       number = {2},
          eid = {157},
        pages = {157},
          doi = {10.3847/1538-4357/acea5a},
archivePrefix = {arXiv},
       eprint = {2302.04291},
 primaryClass = {astro-ph.GA},
       adsurl = {https://ui.adsabs.harvard.edu/abs/2023ApJ...954..157S}
}

@ARTICLE{Juodzbalis2025,
       author = {{Juod{\v{z}}balis}, Ignas and {Maiolino}, Roberto and {Baker}, William M. and {Curtis Lake}, Emma and {Scholtz}, Jan and {D'Eugenio}, Francesco and {Trefoloni}, Bartolomeo and {Isobe}, Yuki and {Tacchella}, Sandro and {Bunker}, Andrew J. and {Carniani}, Stefano and {Charlot}, St{\'e}phane and {Jones}, Gareth C. and {Parlanti}, Eleonora and {Perna}, Michele and {Rinaldi}, Pierluigi and {Robertson}, Brant and {{\"U}bler}, Hannah and {Venturi}, Giacomo and {Willott}, Chris},
        title = "{JADES: comprehensive census of broad-line AGN from Reionization to Cosmic Noon revealed by JWST}",
      journal = {arXiv e-prints},
         year = 2025,
        month = apr,
          eid = {arXiv:2504.03551},
        pages = {arXiv:2504.03551},
archivePrefix = {arXiv},
       eprint = {2504.03551},
 primaryClass = {astro-ph.GA}
}

@ARTICLE{Luo2017,
       author = {{Luo}, B. and {Brandt}, W.~N. and {Xue}, Y.~Q. and {Lehmer}, B. and {Alexander}, D.~M. and {Bauer}, F.~E. and {Vito}, F. and {Ptak}, A. and {Antoniou}, V. and {Ariel}, A. and {Barcons}, A. and {Bellm}, A.~W. and {Boczanowski}, M. and {Braito}, V. and {Chatzopoulos}, I. and {Comastri}, A. and {Gilli}, R. and {Gu}, Q. and {Hornschemeier}, A.~E. and {Iwasawa}, K. and {Johnson}, O. and {Mainieri}, V. and {Ni}, Q. and {Paolillo}, M. and {Rafferty}, D.~A. and {Rosati}, P. and {Shemmer}, O. and {Silverman}, J.~D. and {Tozzi}, P. and {Treister}, E. and {Wang}, J. -X. and {Weaver}, K.~A. and {Xie}, Y. and {Zappacosta}, L.},
        title = "{The Chandra Deep Field-South Survey: 7 Ms Source Catalogs}",
      journal = {\apjs},
         year = 2017,
        month = feb,
       volume = {228},
       number = {1},
          eid = {2},
        pages = {2},
          doi = {10.3847/1538-4365/228/1/2},
archivePrefix = {arXiv},
       eprint = {1611.03501},
 primaryClass = {astro-ph.GA}
}

@article{Antonucci1993,
  author  = {{Antonucci}, Robert},
  title   = "{Unified Models for Active Galactic Nuclei and Quasars}",
  journal = {Annual Review of Astronomy and Astrophysics},
  year    = {1993},
  volume  = {31},
  pages   = {473-521},
  doi     = {10.1146/annurev.aa.31.090193.002353}
}

@article{Urry1995,
  author  = {{Urry}, C. Megan and {Padovani}, Paolo},
  title   = "{Unified Schemes for Radio-Loud Active Galactic Nuclei}",
  journal = {Publications of the Astronomical Society of the Pacific},
  year    = {1995},
  volume  = {107},
  pages   = {803},
  doi     = {10.1086/133630}
}

@article{Netzer2015,
  author  = {{Netzer}, Hagai},
  title   = "{Revisiting the Unified Model of Active Galactic Nuclei}",
  journal = {Annual Review of Astronomy and Astrophysics},
  year    = {2015},
  volume  = {53},
  pages   = {365-408},
  doi     = {10.1146/annurev-astro-082214-122302}
}

@article{Sandles2024,
       author = {{Sandles}, Lester and {D'Eugenio}, Francesco and {Maiolino}, Roberto and {Baker}, William M. and {Bunker}, Andrew J. and {Carniani}, Stefano and {Charlot}, St{\'e}phane and {Curtis-Lake}, Emma and {Hausenius}, C. and {Lo Faro}, Barbara and {Perna}, M. and {Scholtz}, Jan and {Tacchella}, Sandro and {{\"U}bler}, Hannah and {Willott}, Chris J. and {Witstok}, Joris},
        title = "{JADES: The dust attenuation curve of star-forming galaxies at z \raisebox{-0.5ex}\textasciitilde 3}",
      journal = {\aap},
         year = 2024,
        month = dec,
       volume = {691},
          eid = {A305},
        pages = {A305},
          doi = {10.1051/0004-6361/202449622},
archivePrefix = {arXiv},
       eprint = {2402.14070},
 primaryClass = {astro-ph.GA}
}

@misc{Galliano_Galametz_Jones_2018, 
        title={The interstellar dust properties of nearby galaxies | annual reviews}, 
        url={https://www.annualreviews.org/content/journals/10.1146/annurev-astro-081817-051900}, 
        journal={Annual Reviews of Astronomy and Astrophysics}, 
        author={Galliano, Frédéric and Galametz, Maud and Jones, Anthony P},
        year={2018}, 
        month={Jul}
}

@article{Vale_Asari_2020,
   title={Less than the sum of its parts: the dust-corrected H-alpha luminosity of star-forming galaxies explored at different spatial resolutions with MaNGA and MUSE},
   volume={498},
   ISSN={1365-2966},
   url={http://dx.doi.org/10.1093/mnras/staa2557},
   DOI={10.1093/mnras/staa2557},
   number={3},
   journal={Monthly Notices of the Royal Astronomical Society},
   publisher={Oxford University Press (OUP)},
   author={Vale Asari, N and Wild, V and de Amorim, A L and Werle, A and Zheng, Y and Kennicutt, R and Johnson, B D and Galametz, M and Pellegrini, E W and Klessen, R S and Reissl, S and Glover, S C O and Rahner, D},
   year={2020},
   month=Aug, pages={4205–4221} }

@article{Kirkpatrick2012,
       author = {{Kirkpatrick}, Allison and {Dickinson}, Mark and {Chary}, Ranga-Ram and {Mullaney}, James R. and {Pope}, Alexandra and {Alexander}, David M. and {Elbaz}, David and {Fox}, Ori and {Ghosh}, Dhruba and {Good}, Jared and {Goulding}, Andy and {Juneau}, Shardha and {Kartaltepe}, Jeyhan and {Magdis}, Georgios and {Morrison}, Glenn and {Rujopakarn}, Wiphu},
        title = "{GOODS-Herschel: Impact of Active Galactic Nuclei and Star Formation on Dust Heating in Luminous Infrared Galaxies at High Redshift}",
      journal = {\apj},
         year = 2012,
        month = nov,
       volume = {759},
       number = {2},
          eid = {139},
        pages = {139},
          doi = {10.1088/0004-637X/759/2/139},
archivePrefix = {arXiv},
       eprint = {1209.0763},
 primaryClass = {astro-ph.CO}
}

@article{Kennicutt1998,
  author  = {Kennicutt, R. C.},
  title   = {Star formation in galaxies along the Hubble sequence},
  journal = {Annual Review of Astronomy and Astrophysics},
  year    = {1998},
  volume  = {36},
  pages   = {189--232}
}

@article{Murphy2011,
  author  = {Murphy, E. J. and Condon, J. J. and Schinnerer, E. and Kennicutt, R. C. and Calzetti, D. and Armus, L. and Helou, G. and Turner, J. L. and Aniano, G. and Beir{\~a}o, P. and Bolatto, A. D. and Brandl, B. R. and Croxall, K. V. and Dale, D. A. and Meyer, J. L. Donovan and Draine, B. T. and Engelbracht, C. and Hunt, L. K. and Hao, C.-N. and Koda, J. and Roussel, H. and Skibba, R. and Smith, J.-D. T.},
  title   = {CALIBRATING EXTINCTION-FREE STAR FORMATION RATE DIAGNOSTICS WITH 33 GHz FREE-FREE EMISSION IN NGC 6946},
  journal = {The Astrophysical Journal},
  year    = {2011},
  volume  = {737},
  pages   = {67},
  doi     = {10.1088/0004-637x/737/2/67}
}

@ARTICLE{Xue2016,
       author = {{Xue}, Y.~Q. and {Luo}, B. and {Brandt}, W.~N. and {Alexander}, D.~M. and {Bauer}, F.~E. and {Lehmer}, B.~D. and {Yang}, G. -W.},
        title = "{The 2 Ms Chandra Deep Field-North Survey and the 250 ks Extended Chandra Deep Field-South Survey: Improved Source Catalogs}",
      journal = {\apjs},
         year = 2016,
        month = jun,
       volume = {224},
       number = {2},
          eid = {15},
        pages = {15},
          doi = {10.3847/0067-0049/224/2/15},
archivePrefix = {arXiv},
       eprint = {1604.01526},
 primaryClass = {astro-ph.HE}
}

@ARTICLE{Veilleux1995,
       author = {{Veilleux}, S. and {Kim}, D. -C. and {Sanders}, D. B. and {Mazzarella}, J. M. and {Soifer}, B. T.},
        title = "{Optical Spectroscopy of Luminous Infrared Galaxies. II. Analysis of the H II Region and AGN Components}",
      journal = {\apjs},
         year = 1995,
        month = jul,
       volume = {98},
        pages = {171},
          doi = {10.1086/192158},
       adsurl = {https://ui.adsabs.harvard.edu/abs/1995ApJS...98..171V}
}

@article{Calzetti2007,
       author = {{Calzetti}, Daniela and {Kennicutt}, Robert C. and {Engelbracht}, Charles W. and {Leitherer}, Claus and {Draine}, Bruce T. and {Kewley}, Lisa and {Moustakas}, John and {Sosey}, Michael and {Dale}, Daniel A. and {Gordon}, Karl D. and {Helou}, George X. and {Hollenbach}, David J. and {Armus}, Lee and {Bendo}, George J. and {Bot}, Caroline and {Buckalew}, Brent and {Jarrett}, Thomas and {Li}, Aigen and {Meyer}, Martin and {Murphy}, Eric J. and {Prescott}, Moire and {Regan}, Michael W. and {Rieke}, George H. and {Roussel}, H{\'e}l{\`e}ne and {Sheth}, Kartik and {Smith}, J. D. T. and {Thornley}, Michele D. and {Walter}, Fabian},
        title = "{The Calibration of Mid-Infrared Star Formation Rate Indicators}",
      journal = {\apj},
         year = 2007,
        month = dec,
       volume = {666},
       number = {2},
        pages = {870-895},
          doi = {10.1086/520082},
archivePrefix = {arXiv},
       eprint = {0705.3414},
       adsurl = {https://ui.adsabs.harvard.edu/abs/2007ApJ...666..870C}
}

@ARTICLE{2022A&A...661A..80J,
       author = {{Jakobsen}, P. and {Ferruit}, P. and {Alves de Oliveira}, C. and {Arribas}, S. and {Bariani}, M. and {Birkmann}, S.~M. and {B{\"o}ker}, T. and {Bunker}, A.~J. and {Caballero}, J.~A. and {Charlot}, S. and {de Marchi}, G. and {Franx}, M. and {Giardino}, G. and {Glowacki}, R. and {Husemann}, B. and {Jakobsen}, P. and {Korneef}, J. and {L{\"u}tzgendorf}, N. and {Maiolino}, R. and {Manjavacas}, E. and {Marston}, A. and {Moseley}, S.~H. and {Phaff}, J. and {Pirzkal}, N. and {Rawle}, T.~D. and {Rauscher}, B.~J. and {Rix}, H. -W. and {Sabbi}, E. and {Sirianni}, M. and {te Plate}, M. and {Valenti}, J. and {Willott}, C.~J. and {Zeidler}, P.},
        title = "{The Near-Infrared Spectrograph (NIRSpec) on the James Webb Space Telescope. I. Overview of the instrument and its capabilities}",
      journal = {\aap},
         year = 2022,
        month = may,
       volume = {661},
          eid = {A80},
        pages = {A80},
          doi = {10.1051/0004-6361/202243360},
archivePrefix = {arXiv},
       eprint = {2202.03305},
 primaryClass = {astro-ph.IM},
       adsurl = {https://ui.adsabs.harvard.edu/abs/2022A&A...661A..80J}
}

@article{2013A&A...558A..33A,
       author = {{Astropy Collaboration} and {Robitaille}, Thomas P. and {Tollerud}, Erik J. and {Greenfield}, Perry and {Droettboom}, Michael and {Bray}, Erik and {Aldcroft}, Tom and {Davis}, Matt and {Ginsburg}, Adam and {Price-Whelan}, Adrian M. and {Kerzendorf}, Wolfgang E. and {Conley}, Alex and {Crighton}, Neil and {Barbary}, Kyle and {Muna}, Demitri and {Ferguson}, Henry and {Grollier}, Fr{\'e}d{\'e}ric and {Parikh}, Madhura M. and {Nair}, Prasanth H. and {Unther}, Hans M. and {Deil}, Christoph and {Woillez}, Julien and {Conseil}, Simon and {Kramer}, Roban and {Wheeler}, J. E. H. and {Tavrov}, Alexey and {Siegel}, Braden and {Sip{\H{o}}cz}, Brigitta M. and {Patil}, Manodeep and {Garg}, Suyog and {Gondhalekar}, Yash and {Lian}, T. Y. and {Lim}, Pey Lian and {Bostroem}, K. Azalee and {G{\"u}nther}, H. Moritz and {Kelley}, Michael S. P. and {Barentsen}, Geert and {Pascual}, Sergio and {Souchereau}, G. and {Williams}, B. and {Lewenza}, S. and {Lust}, A. and {Lian}, K. and {PASP}, T. A. S. O. T.},
        title = "{Astropy: A community Python package for astronomy}",
      journal = {\aap},
         year = 2013,
        month = oct,
       volume = {558},
          eid = {A33},
        pages = {A33},
          doi = {10.1051/0004-6361/201322068},
archivePrefix = {arXiv},
       eprint = {1307.6212},
 primaryClass = {astro-ph.IM},
       adsurl = {https://ui.adsabs.harvard.edu/abs/2013A&A...558A..33A}
}

@article{2018AJ....156..123A,
       author = {{Astropy Collaboration} and {Price-Whelan}, Adrian M. and {Sip{\H{o}}cz}, Brigitta M. and {G{\"u}nther}, Hans Moritz and {Lim}, Pey Lian and {Crawford}, Stuart M. and {Conseil}, Simon and {Shupe}, David L. and {Craig}, Matthew W. and {Dencheva}, Nadia and {Ginsburg}, Adam and {Vand erPlas}, Jake T. and {Bradley}, Larry D. and {P{\'e}rez-Su{\'a}rez}, David and {de Val-Borro}, Miguel and {Aldcroft}, Thomas L. and {Cruz}, Kelle L. and {Robitaille}, Thomas P. and {Tollerud}, Erik J. and {Ardelean}, Crispin and {Babej}, Tom and {Bachetti}, Matteo and {Barentsen}, Geert and {Bostroem}, K. Azalee and {Droettboom}, Michael and {Ferreira}, Leonardo and {Garland}, Ryan and {Ginsburg}, Clara and {Goodman}, Alyssa A. and {Greco}, Jennifer and {Greenfield}, Perry and {Groener}, Austen M. and {Hagen}, Alex and {Hirst}, Paul and {Homeier}, Derek and {Horton}, Anthony J. and {Hosseinzadeh}, Griffin and {Hu}, Lei and {Hunkeler}, J. S. and {Ivezi{\'c}}, {\v{Z}}eljko and {Kervella}, Pierre and {Lian}, K. and {Lian}, T. Y. and {Lust}, A. and {Ma}, Zhiyuan and {MacBride}, Conor and {Marocco}, Federico and {Maljaars}, Jakob and {Muna}, Demitri and {Murphy}, N. A. and {Norman}, Henrik and {O'Steen}, Richard and {Oman}, Kyle A. and {Pacifici}, Camilla and {Pascual}, Sergio and {Pascual-Granado}, J. and {Patil}, Aarya A. and {Perren}, Gabriel I. and {Pickering}, Timothy E. and {Pintado}, O. and {Popov}, S. and {Rector}, T. A. and {Renzini}, A. and {Reyes}, J. and {Richards}, J. W. and {Saba}, A. and {Samos-S{\'a}enz de Tejada}, A. and {Sanders}, J. and {Pandey}, P. and {Lian}, K. and {Lian}, K. and {PASP}, T. A. S. O. T.},
        title = "{The Astropy Project: Building an Open-science Project and Status of the v2.0 Core Package}",
      journal = {\aj},
         year = 2018,
        month = sep,
       volume = {156},
       number = {3},
          eid = {123},
        pages = {123},
          doi = {10.3847/1538-3881/aabc4f},
archivePrefix = {arXiv},
       eprint = {1801.02634},
 primaryClass = {astro-ph.IM},
       adsurl = {https://ui.adsabs.harvard.edu/abs/2018AJ....156..123A}
}

@article{2022ApJ...935..167A,
       author = {{Astropy Collaboration} and {Price-Whelan}, Adrian M. and {Lim}, Pey Lian and {Earl}, Nicholas and {Starkman}, Nathaniel and {Bradley}, Larry and {Shupe}, David L. and {Patil}, Aarya A. and {Corrales}, Lia and {Brasseur}, C. E. and {N{\"o}the}, Maximilian and {Donath}, Axel and {Tollerud}, Erik and {Morris}, Brett M. and {Ginsburg}, Adam and {Vaher}, Eero and {Weaver}, Benjamin A. and {Tocknell}, James and {Jamieson}, William and {van Kerkwijk}, Marten H. and {Robitaille}, Thomas P. and {Merry}, Bruce and {Bachetti}, Matteo and {G{\"u}nther}, H. Moritz and {Aldcroft}, Thomas L. and {Alvarado-Montes}, Jaime A. and {Archibald}, Anne M. and {B{\'o}di}, Attila and {Bapat}, Shreyas and {Barentsen}, Geert and {Baz{\'a}n}, Juanjo and {Biswas}, Manish and {Boquien}, M{\'e}d{\'e}ric and {Burke}, D. J. and {Cara}, Daria and {Cara}, Mihai and {Conroy}, Kyle E. and {Conseil}, Simon and {Craig}, Matthew W. and {Cross}, Robert M. and {Cruz}, Kelle L. and {D'Eugenio}, Francesco and {Dencheva}, Nadia and {Devillepoix}, Hadrien A. R. and {Dietrich}, J{\"o}rg P. and {Eigenbrot}, Arthur Davis and {Erben}, Thomas and {Ferreira}, Leonardo and {Foreman-Mackey}, Daniel and {Fox}, Ryan and {Freij}, Nabil and {Garg}, Suyog and {Geda}, Robel and {Glattly}, Lauren and {Gondhalekar}, Yash and {Gordon}, Karl D. and {Grant}, David and {Greenfield}, Perry and {Groener}, Austen M. and {Guest}, Steve and {Gurovich}, Sebastian and {Handberg}, Rasmus and {Hart}, Akeem and {Hatfield-Dodds}, Z. and {Homeier}, Derek and {Hosseinzadeh}, Griffin and {Jenness}, Tim and {Jones}, Craig K. and {Joseph}, Prajwel and {Kalmbach}, J. Bryce and {Karamehmetoglu}, Emir and {Ka{\l}uszy{\'n}ski}, Miko{\l}aj and {Kelley}, Michael S. P. and {Kern}, Nicholas and {Kerzendorf}, Wolfgang E. and {Koch}, Eric W. and {Kulumani}, Shankar and {Lee}, Antony and {Ly}, Chun and {Ma}, Zhiyuan and {MacBride}, Conor and {Maljaars}, Jakob M. and {Muna}, Demitri and {Murphy}, N. A. and {Norman}, Henrik and {O'Steen}, Richard and {Oman}, Kyle A. and {Pacifici}, Camilla and {Pascual}, Sergio and {Pascual-Granado}, J. and {Patil}, Rohit R. and {Perren}, Gabriel I. and {Pickering}, Timothy E. and {Rastogi}, Tanuj and {Roulston}, Benjamin R. and {Ryan}, Daniel F. and {Rykoff}, Eli S. and {Sabater}, Jose and {Sakurikar}, Parikshit and {Salgado}, Jes{\'u}s and {Sanghi}, Aniket and {Saunders}, Nicholas and {Savchenko}, Volodymyr and {Schwardt}, Ludwig and {Seifert-Eckert}, Michael and {Shankar}, R. and {Sharma}, M. and {Sherbert}, L. and {Short}, P. and {Sienkiewicz}, F. and {Sillero}, E. and {Singer}, Leo P. and {Singrol}, A. and {Sinha}, M. and {Sip{\H{o}}cz}, Brigitta M. and {Spitler}, Lee and {Stansby}, D. and {Streicher}, O. and {Subramani}, K. and {S{\"u}veges}, M{\'a}ria and {Swinbank}, J. and {Tey}, J. S. and {Teague}, R. and {Tiwari}, P. and {de Val-Borro}, Miguel and {Vemparala}, N. and {Wargelin}, B. and {Warren}, A. R. and {Warwick}, S. and {Weir}, D. J. and {WeQT}, R. and {Wenger}, T. V. and {Wiebe}, N. and {Wootten}, A. and {Wright}, D. and {YAP}, K. T. and {Yoachim}, P. and {Zhang}, C. and {Zonca}, A. and {Astropy Project}, T.},
        title = "{The Astropy Project: Sustaining and Growing a Community-oriented Open-source Project and Package}",
      journal = {\apj},
         year = 2022,
        month = aug,
       volume = {935},
       number = {2},
          eid = {167},
        pages = {167},
          doi = {10.3847/1538-4357/ac7c74},
archivePrefix = {arXiv},
       eprint = {2201.07184},
 primaryClass = {astro-ph.IM},
       adsurl = {https://ui.adsabs.harvard.edu/abs/2022ApJ...935..167A}
}

@article{harris2020array,
       author = {{Harris}, Charles R. and {Millman}, K. Jarrod and {van der Walt}, St{\'e}fan J. and {Gommers}, Ralf and {Virtanen}, Pauli and {Cournapeau}, David and {Wieser}, Eric and {Taylor}, Julian and {Berg}, Sebastian and {Smith}, Nathaniel J. and {Kern}, Robert and {Picus}, Matti and {Hoyer}, Stephan and {van Kerkwijk}, Marten H. and {Brett}, Matthew and {Haldane}, Allan and {del R{\'\i}o}, Jaime Fern{\'a}ndez and {Wiebe}, Mark and {Peterson}, Pearu and {G{\'e}rard-Marchant}, Pierre and {Sheppard}, Kevin and {Reddy}, Tyler and {Weckesser}, Warren and {Gabbard}, Hameer and {Terasvouri}, Miika and {Zaide}, Ilhan and {Javanmard}, Pavan and {H{\"a}fner}, Simon and {Aroh}, Travis and {Nelson}, Andrew and {Taskesen}, Ahmet and {SciPy 1.0 Contributors}},
        title = "{Array programming with NumPy}",
      journal = {\nat},
         year = 2020,
        month = sep,
       volume = {585},
       number = {7825},
        pages = {357-362},
          doi = {10.1038/s41586-020-2649-2},
archivePrefix = {arXiv},
       eprint = {2006.10256},
 primaryClass = {cs.MS},
       adsurl = {https://ui.adsabs.harvard.edu/abs/2020Natur.585..357H}
}

@article{2020SciPy-NMeth,
  author  = {Virtanen, Pauli and Gommers, Ralf and Oliphant, Travis E. and
            Haberland, Matt and Reddy, Tyler and Cournapeau, David and
            Burovski, Evgeni and Peterson, Pearu and Weckesser, Warren and
            Bright, Jonathan and {van der Walt}, St{\'e}fan J. and
            Brett, Matthew and Wilson, Joshua and Millman, K. Jarrod and
            Mayorov, Nikolay and Nelson, Andrew R. J. and Jones, Eric and
            Kern, Robert and Larson, Eric and Carey, C J and
            Polat, {\.I}lhan and Feng, Yu and Moore, Eric W. and
            {VanderPlas}, Jake and Laxalde, Denis and Perktold, Josef and
            Cimrman, Robert and Henriksen, Ian and Quintero, E. A. and
            Harris, Charles R. and Archibald, Anne M. and
            Ribeiro, Ant{\^o}nio H. and Pedregosa, Fabian and
            {van Mulbregt}, Paul and {SciPy 1.0 Contributors}},
  title   = {{{SciPy} 1.0: Fundamental Algorithms for Scientific
            Computing in Python}},
  journal = {Nature Methods},
  year    = {2020},
  volume  = {17},
  pages   = {261--272},
  adsurl  = {https://rdcu.be/b08Wh},
  doi     = {10.1038/s41592-019-0686-2},
}

@article{Hunter:2007,
  Author    = {Hunter, J. D.},
  Title     = {Matplotlib: A 2D graphics environment},
  Journal   = {Computing in Science \& Engineering},
  Volume    = {9},
  Number    = {3},
  Pages     = {90--95},
  publisher = {IEEE COMPUTER SOC},
  doi       = {10.1109/MCSE.2007.55},
  year      = 2007
}

\end{document}